\documentclass[conference]{IEEEtran}
\IEEEoverridecommandlockouts
\usepackage{cite}
\usepackage{amsmath,amssymb,amsfonts}
\usepackage{algorithm, algpseudocode}
\usepackage{graphicx}
\usepackage{textcomp}
\usepackage{xcolor}
\def\BibTeX{{\rm B\kern-.05em{\sc i\kern-.025em b}\kern-.08em
    T\kern-.1667em\lower.7ex\hbox{E}\kern-.125emX}}

\newtheorem{theorem}{Theorem}

\newtheorem{lemma}{Lemma}

\def\qed{\hbox{${\vcenter{\vbox{			
   \hrule height 0.4pt\hbox{\vrule width 0.4pt height 6pt
   \kern5pt\vrule width 0.4pt}\hrule height 0.4pt}}}$}}

\begin{document}

\title{Minimizing Intellectual Property Risks via Self-Stabilizing Algorithms}

\author{\IEEEauthorblockN{2\textsuperscript{nd} Iman Evazzade}
\IEEEauthorblockA{\textit{Information Technology Research Center} \\
\textit{BMW Group}\\
Greenville, SC \\
Iman.Evazzade@bmwgroup.com}
\and
\IEEEauthorblockN{1\textsuperscript{st} Ken Kennedy}
\IEEEauthorblockA{\textit{Information Technology Research Center} \\
\textit{BMW Group}\\
Greenville, SC \\
Ken.Kennedy@bmwgroup.com}
}

\maketitle

\begin{abstract}

In this paper, we examine the use of self-stabilizing algorithms, operating in a hierarchical manner, to determine intellectual property risks at a macro level. We are both interested in finding a solution that will support all defined intellectual property dimensions as well as suboptimal solutions in order to minimize risk.

\end{abstract}

\section{Introduction}

This paper focuses on finding, and reducing, intellectual property risks at a macro level. We originally sought to find prior work that could help solve this problem. The solution proposed by O'Kane and Shell, which used graph coloring, partially met our requirements for determining risk \cite{7139023}.

\begin{figure}[htbp]
  \centerline{\includegraphics[scale=0.38, angle=-90]{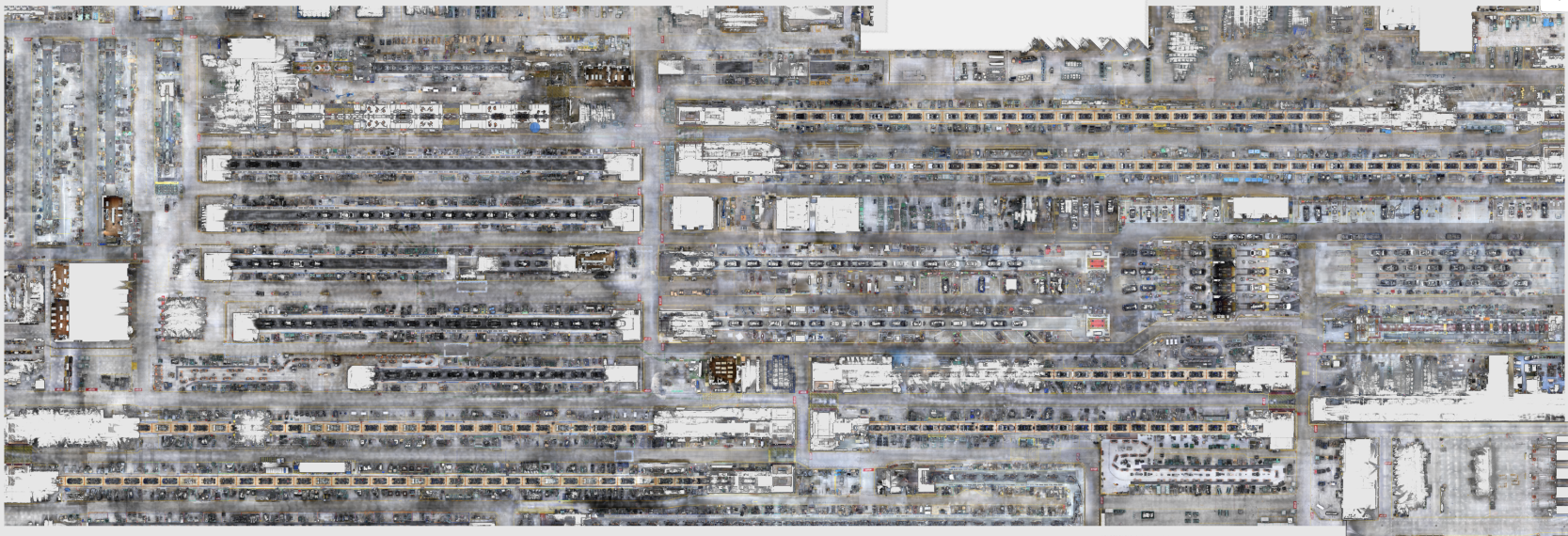}}
  \caption{Overview of hall.}
  \label{fig:hall_overview}
\end{figure}

However, their solution was NP-hard and would only work for toy problems in which a brute force solution could also easily be used. In our case, we required a solution that could be deployed at industrial scale for determining risk within, and across, manufacturing halls. A top-down view of one such hall is shown in Figure~\ref{fig:hall_overview}.

In Section~\ref{sec:background}, we present prior work from different groups that have examined both this problem and intellectual property protection in general. Section~\ref{sec:ip_dimensions} describes a subset of the intellectual property dimensions that we would like to examine. Our approach is presented in Section~\ref{sec:algorithms}, and conclusions are provided in Section~\ref{sec:conclusions}.

\section{Background}\label{sec:background}

The motivation for this paper came from \cite{7139023}, which uses graph coloring for privacy. A similar paper from some of the same authors examines the topic of planning and privacy with robots \cite{10.1007-978-3-030-44051-0-7}. While not directly examining IP within an area, Zare-Garizy et. al. examine privacy topics within the supply chain \cite{10.1155-2018-3858592}.

A few general anonymization techniques of data include 
\textit{k}-anonymity \cite{10.1142/S0218488502001648}, 
${\ell}$-diversity \cite{10.1145/1217299.1217302}, 
\textit{t}-Closeness \cite{4221659}, 
differential privacy \cite{10.1007/11787006_1}, and 
differential private analysis of graphs \cite{Raskhodnikova2014}. An additional method of protecting intellectual property includes obfuscation techniques \cite{7004228}. These have been used extensively for helping to protect various forms of intellectual property. Some of these techniques would apply to minimizing IP risks at a micro level. For instance, protecting information at a certain location within a manufacturing hall.

Extensions to the techniques above for graph anonymization include edge differential privacy of graphs\cite{Raskhodnikova2016}, \cite{DBLP:journals/corr/abs-2010-08688} and node differential privacy \cite{DBLP:journals/corr/abs-1204-2136}, \cite{10.1007-978-3-642-36594-2-26}. Our representation of IP risks will use graphs and some of these techniques could have some applications.

\section{Intellectual Property Dimensions}\label{sec:ip_dimensions}

In this section, we discuss some of the intellectual property that we may are interested in protecting. This is not an exhaustive list but is drawn from the focus of the project that we were examining at the time.

\subsection{White / Blacklist Columns}

When discussing a white / blacklist for a column, a column is a physical location within a manufacturing hall. This can be viewed the same as a chess board in which column I13 will uniquely refer to a spot, and its immediate vicinity. For some of the examples below, we will exclude the numeric component of a column for brevity.

For whitelists, there is no activity at that location that needs to be safeguarded while a blacklisted column has intellectual property that we do not want to be observed. Examples of this for our use case included scrap and rework. However, there could be other cases, such as achievement / monitoring boards visible that contain sensitive information regarding how the hall is functioning.

\subsection{Associates / Equipment / Suppliers}

An additional vector for IP and privacy included information about associates, equipment, and suppliers. For instance, we might not mind if a subset of these are known, but we would not like for all of them to be so.

A generic example regarding the type of information that might be available within a manufacturing hall, at specific columns, is shown in Table~\ref{tab:companies}.

\begin{table}
    \centering
    \begin{tabular}{c|c}
        \textbf{Column} & \textbf{Supplier} \\
        \hline
        A & X\\
        \hline
        B & X, Y\\
        \hline
        C & Y\\
        \hline
        D & Y, Z\\
        \hline
        E & Z\\
        \hline
        F & X, Z\\
        \hline
    \end{tabular}
    \caption{Simple example of associates from different companies visible at different column locations.}
    \label{tab:companies}
\end{table}

There are many different methods that could be used here to ensure that all of the suppliers are not visible. For example, we could manually blacklist any columns that contain a certain supplier. However, we choose to use a maximal independent set here as it would both guarantee that not all information would be seen and we did not have a list ahead of time of which suppliers should, or should not, be seen.

\subsection{Parts Flow}

\begin{figure}[htbp]
  \centerline{\includegraphics[scale=0.8]{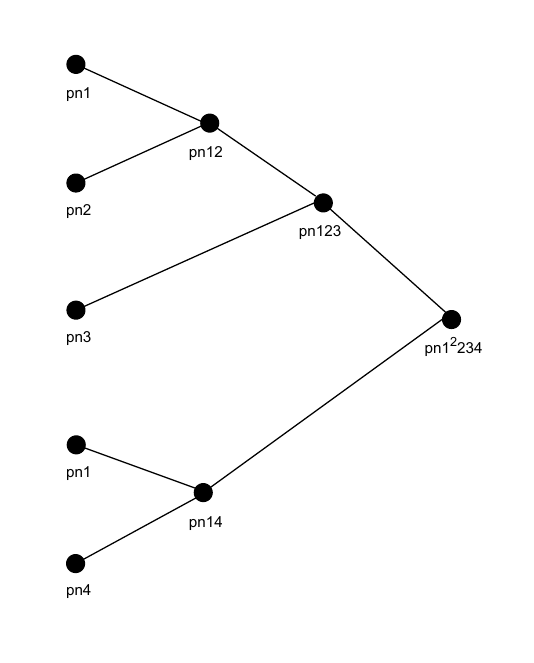}}
  \caption{Example of parts flow.}
  \label{fig:parts_flow}
\end{figure}

During the manufacturing process, we are starting with a large number of parts that are continuously being assembled into one final product. A simple example of this flow is shown in Figure~\ref{fig:parts_flow}.

For privacy, it may be desirable to exclude some portion of this process. This is similar to the supplier use case above. As such, we could use whitelists/blacklists, but we do not have particular products to include/exclude. For this case, we use a 1-minimal dominating set, but a maximal independent set could also have been used. A coloring of the graph would also have been appropriate.

\section{Algorithms}\label{sec:algorithms}

We follow the general approach in \cite{7139023}; however, we are interested in applying the algorithm to a real-world use case. The algorithm by O'Kane and Shell is NP-hard and not suitable for large scale problems. In their paper, O'Kane and Shell were focused on finding $\chi(G)$, the chromatic number of the graph. We will instead focus on minimal/maximal sets rather than minimum/maximum in order to move the complexity for subproblems to polynomial time.

First, we provide a few definitions. Let us use a graph $G = (V, E)$ that is connected with order $n = |V|$ and $m = |E|$. The open neighborhood of a node $i \in V$ is $N(i) = {j | ij \in E}$, and the closed neighborhood of node $i \in V$ is $N[i] = N(i) \cup {i}$.

A set $S$ is \textit{dominating set} if $N[S] = V$, where $v \in V$ is in \textit{S} or adjacent to a node in \textit{S}. A minimal dominating set is one in which if any node in $S$ is removed, the set is no longer dominating. The minimum cardinality of a dominating set is $\gamma$ and the maximum cardinality is $\Gamma$.

A set $S$ is independent if no two nodes in $S$ are adjacent. A maximal independent set is one where if any node in $V \backslash S$ is added,  then $S$ is no longer independent. The minimum cardinality of a maximal independent set is $i(G)$ and $\beta_0$ is the maximum cardinality.

We also define here an irredundant set, which is one for which there is not a $v \in V$ in which $N[S - {v}] \neq N[S]$. The lower irredundance number of a graph $G$ is $ir$, and $IR$ is the upper irredundance number. The following inequality comes from \cite{cockayne1987k}.

\begin{equation}
    ir \leq \gamma \leq i \leq \beta_0 \leq \Gamma \leq IR
    \label{eq:inequality}
\end{equation}

\subsection{Non-convergence of Simultaneous Solutions with Equal Priority}\label{subsec:non_convergence}

For a single graph, a maximal independent set is also a minimal dominating set \cite{haynes2013fundamentals}. However in our case, we are dealing with multiple graphs that share Columns (nodes or subnodes) but may have a different structure.

Here we will walk through an example using a maximal independent set to represent suppliers and a 1-minimal dominating set for parts flow in order to see a case in which solving the graphs simultaneously with equal priority of IP dimensions will not lead to convergence.

\begin{figure}[htbp]
  \centerline{\includegraphics[page=1, scale=0.5]{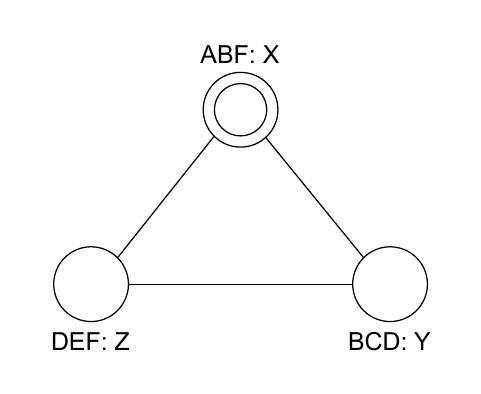}}
  \caption{Compacted graph with maximal independent set from Table I.}
  \label{fig:ind_set_example}
\end{figure}

Using the nodes and suppliers defined in Table~\ref{tab:companies}, we arrive at Figure~\ref{fig:ind_set_example} with one node in the independent set. Note that in this case, the graph is compacted as the full version would have a clique for each node. This graph can also be built different ways depending upon what information is deemed acceptable to share. For instance, it may be that we do not mind if $Y$ is known. In this case, the $BCD$ node would not have any connections to the other nodes.

\begin{figure}[htbp]
  \centerline{\includegraphics[page=1, scale=0.4]{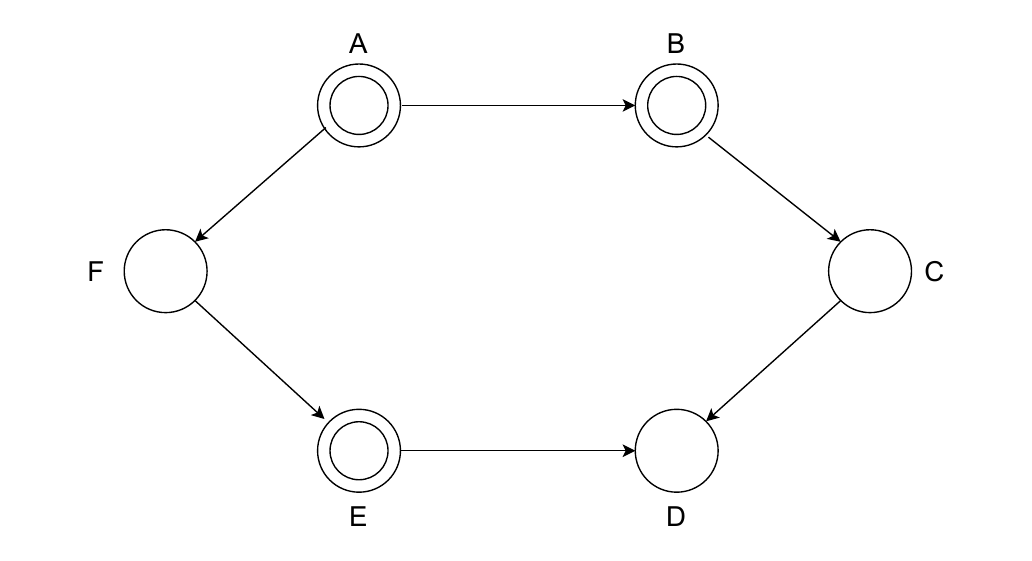}}
  \caption{Simple example of a minimal dominating set.}
  \label{fig:dom_set_example}
\end{figure}

Let us then create a parts flow graph for the columns in Table~\ref{tab:companies} and assign a minimal dominating - one such set is shown in Figure~\ref{fig:dom_set_example}. Note that the flow we have created is depicted via directional edges.

Any dominating set of this graph will result in at least two adjacent nodes of Figure~\ref{fig:ind_set_example} being put into the independent set; additionally, any independent set of Figure~\ref{fig:ind_set_example} will not be a dominating set in Figure~\ref{fig:dom_set_example}. Thus trying to apply the constraints over both of the graphs simultaneously will never converge as the independent and dominating sets are in contention.

\subsection{Methodology}

In order to solve for the graphs, we can either attempt to do so sequentially or simultaneously in which IP elements do not have equal priority. If the latter, there are different strategies we may use that include cellular automata \cite{neumann1966theory} and self-stabilization \cite{dijkstra1974self} \cite{dijkstra1985belated}. For this paper, we will be using the self-stabilizing paradigm with an unfair distributed scheduler.

There are a multitude of algorithms that could be used for IP protection. In this paper, we will focus on just the three use cases and the algorithms mentioned for them: blacklists/whitelists, maximal independent sets, and 1-minimal dominating sets. For each of these, we will need to show both correctness and convergence. Note that since there will be a priority of how the algorithms are applied, correctness (e.g. a maximal independent set) will be on the subgraph $V_1 \subseteq V$ induced from $V$ by an algorithm with higher priority rather than on $V$ itself; otherwise, we would have the same issue of convergence that was mentioned above.

\subsection{Algorithm for White and Blacklists}\label{subsec:bw}

An algorithm is included here for determining if a node is within white and/or blacklists. For a node $i \in V$, the variable $BW(i)$ will be used to indicate if the node is within a whitelist or blacklist: $BW(i) = in$ or $BW(i) = out$ respectively.

\begin{algorithm}[htbp]
 \begin{algorithmic}
    \State \textbf{RWait: if} $x_{BW}$(i) = ${out}$ ${\wedge}$ $BW(i)$ = ${in}$ ${\wedge}$ (${\nexists}$a ${\in}$ $ALG$)($a < BW$ ${\wedge}$ $x_{a}$(i) = ${out}$)
    \State ${\;}$ \textbf{then} $x_{BW}$(i) = ${wait}$
    
    \State \textbf{RBack: if} $x_{BW}$(i) = ${wait}$ ${\wedge}$ ($BW(i)$ $\neq$ ${in}$ $\vee$ (${\exists}$a ${\in}$ $ALG$)($a < BW$ ${\wedge}$ ($x_{a}$(i) = ${out}$)))
    \State ${\;}$ \textbf{then} $x_{BW}$(i) = ${out}$

    \State \textbf{RIn: if} $x_{BW}$(i) = ${wait}$ ${\wedge}$ $BW(i)$ = ${in}$ ${\wedge}$ (${\nexists}$a ${\in}$ $ALG$)($a < BW$ ${\wedge}$ $x_{a}$(i) = ${out}$)
    \State ${\;}$ \textbf{then} $x_{BW}$(i) = ${in}$

    \State \textbf{ROut: if} ($x_{BW}$(i) = ${in}$ $\vee$ $x_{BW}$(i) = ${wait}$) ${\wedge}$ ($BW(i)$ = ${out}$ ${\vee}$ (${\exists}$a ${\in}$ $ALG$)($a < BW$ ${\wedge}$ $x_{a}$(i) = ${out}$))
    \State ${\;}$ \textbf{then} $x_{BW}$(i) = ${out}$
 
 \end{algorithmic}
 \caption{White and Blacklists}
 \label{alg:bw_combined}
\end{algorithm}

We have added a component in which each algorithm will have a numerical id that equates to its priority. For the example above, the algorithm is $BW$. If there were multiple versions of this algorithm in a use case, then you would have $BW_1, BW_2, ...$.

\begin{table}
    \centering
    \begin{tabular}{c|c|c}
        \textbf{Node} & \textbf{$BW_1$} & \textbf{$BW_2$} \\
        \hline
        A & $in$ & $in$\\
        \hline
        B & $in$ & $in$\\
        \hline
        C & $out$ & $in$\\
        \hline
        D & $out$ & $out$\\
        \hline
        E & $in$ & $out$\\
        \hline
    \end{tabular}
    \caption{Multiple white and blacklists across different nodes.}
    \label{tab:mul_bw_example}
\end{table}

\begin{figure}[htbp]
  \centerline{\includegraphics[page=1, scale=0.6]{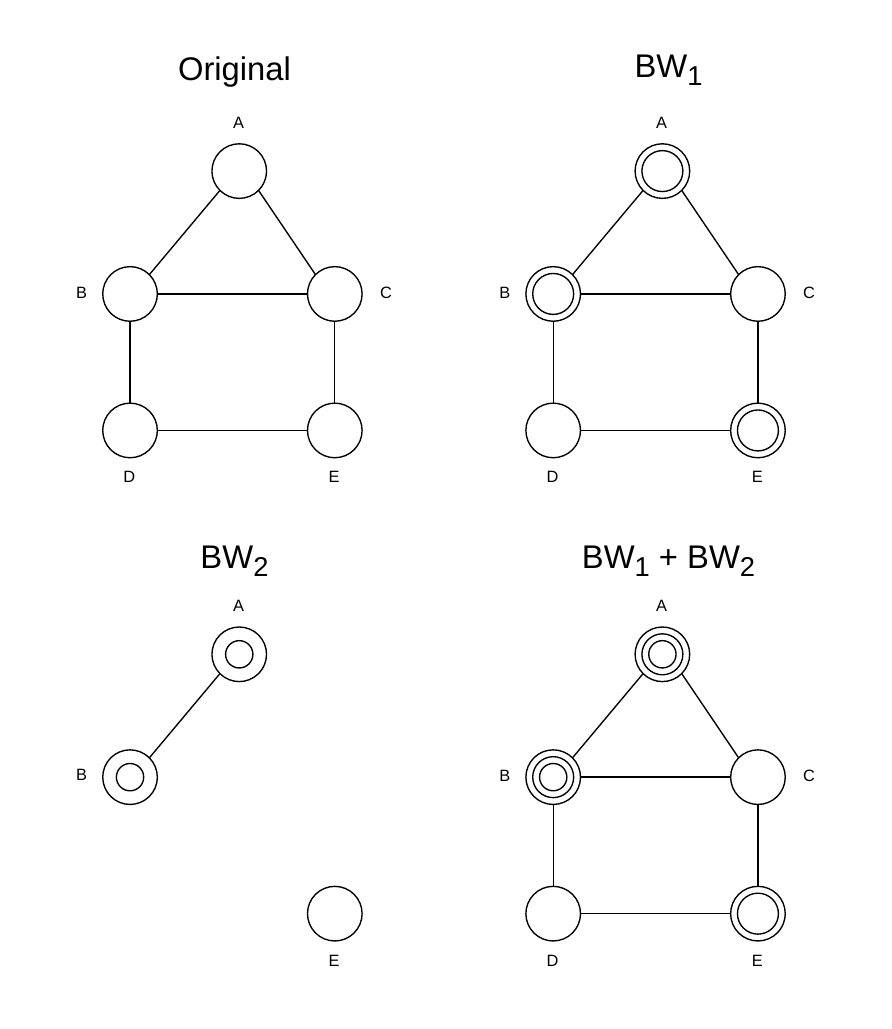}}
  \caption{Graphical representation of Table~\ref{tab:mul_bw_example}. $BW_1$ is applied to the original graph. This creates a subgraph on which $BW_2$ can be applied. The final result is then a combination of both in which only nodes A and B satisfy all of the $in$ criteria.}
  \label{fig:mul_bw_example}
\end{figure}

\lemma{When the system is stable, no node $i$ has $x_{BW}(i) = wait$.}

\textbf{Proof:} Let us assume that the system is stable, and there is a case in which a node $i$ has $x_{BW}(i) = wait$. If $BW(i) \neq in$, then rule \textbf{RBack} would be executed. Thus, we are only considering the case in which $BW(i) = in$. Rules \textbf{RBack} and \textbf{RIn} consider the mutually exclusive cases in which there is, or is not, an algorithm that has a lower id than $BW$ with $x_a(i) = out$. Therefore, either rule \textbf{RBack} or \textbf{RIn} would have been executed and $x_{BW}(i)$ would not have state $wait$ after the system has stabilized. Thus, this is not a case that can happen. \qed

\lemma{When the system is stable, the set $X$ of nodes $x_{BW}(i)$ will have the value $out$ if $BW(i) = out$.}

\textbf{Proof:}  Let us assume that there is a node $i$ in which $BW(i) = out$ but $x_{BW}(i) \neq out$. In this case, rule \textbf{ROut} would be activated, but this contradicts the assumption that the system was stable. \qed

\lemma{When the system is stable, the set $X$ of nodes $x_{BW}(i)$ will have the value of $in$ if $BW(i) = in$ for the subgraph created from algorithms $a \in A$ where $a < BW$.}

\textbf{Proof:} There are two cases to consider. First case is when $BW(i) = in$ for a node $i$ in which ${\exists}$a ${\in}$ $ALG$)($a < BW$ where $x_{a}$(i) = ${out}$. However, this would lead to rule \textbf{ROut} being activated. For the second case, let us consider where $x_{BW}(i) \neq in$ but ${\forall}$a ${\in}$ $ALG$)($a < BW$ ${\wedge}$ $x_{a}$(i) = ${in}$). In this case, either \textbf{RWait} would be activated if $x_{BW}(i) = out$ or \textbf{RIn} would be activated if $x_{BW}(i) = wait$. Therefore, the system was not stable. \qed

\lemma{When the system is stable, the set $X$ of nodes forms a white and blacklist on the subgraph induced by $(a \in ALG)(a < MIS)$.}

\textbf{Proof:} See lemmas 2-3. \qed

\lemma{If there are no more moves for a node $i$ due to an algorithm $a$ in which $a < BW$, then if node $i$ executes \textbf{RIn}, it will never make another move.}

\textbf{Proof:} If a node $i$ executes \textbf{RIn}, then it must be that $BW(i) = in$ and ${\nexists}$a ${\in}$ $ALG$)($a < BW$ ${\wedge}$ $x_{a}$(i) = ${out}$). The only rule that can be executed when $x_{a}$(i) = ${in}$ is \textbf{ROut}. However, this would require that $BW(i) = out$ or ${\exists}$a ${\in}$ $ALG$)($a < BW$ ${\wedge}$ $x_{a}$(i) = ${out}$). As we have already stated that there were no subsequent moves from algorithms with $a < BW$ and $BW(i) = in$. Thus \textbf{ROut} could not be activated and there will not be any further moves on $i$. \qed

\lemma{If there are no more moves for a node $i$ due to an algorithm $a$ in which $a < BW$, then if node $i$ executes \textbf{ROut}, it will never make another move.}

\textbf{Proof:} If this move is executed and $BW(i) = out$, then no other move will be executed. The only other rule that does not require $BW(i) = in$ is \textbf{RBack}, but it requires $x_{BW}(i) = wait$. However, this will be set to $x_{BW}(i) = out$ after \textbf{ROut}, so it will not be activated. Likewise, if this move is executed due to ${\exists}$a ${\in}$ $ALG$)($a < BW$ ${\wedge}$ $x_{a}$(i) = ${out}$), then moves \textbf{RIn} and \textbf{RWait} cannot be executed due to the condition that there are no more moves for a node $i$ due to an algorithm $a$ in which $a < BW$. \qed

\lemma{If there are no more moves for a node $i$ due to an algorithm $a$ in which $a < BW$, then if node $i$ executes \textbf{RBack}, it will never make another move.}

\textbf{Proof:} As the conditions for \textbf{RBack} are mutually exclusive to \textbf{RWait}, and there will be no further changes due to $a < BW$, then \textbf{RWait} will not be activated. Additionally, neither \textbf{RIn} or \textbf{ROut} can be activated as $x_{BW}(i)$ will be set to $out$. \qed

\lemma{If there are no more moves for a node $i$ due to an algorithm $a$ in which $a < BW$, then if node $i$ executes \textbf{RWait}, there will be at most one other move.}

\textbf{Proof:} The potential moves are \textbf{RIn}, \textbf{ROut}, or \textbf{RBack}. From lemmas 5-7, we see that each of these will be the last move made. \qed

\theorem{Algorithm~\ref{alg:bw_combined} finds a white/blacklist in at most $O(2n)$ moves.}

\textbf{Proof:} This follows directly from lemmas 4-8. \qed

\subsection{Algorithm for Maximal Independent Set}\label{subsec:mis}

Initial work on finding a maximal independent set via a self-stabilizing algorithm was done by Shukla, Rosenkrantz and Ravi \cite{shukla}. They assumed a central scheduler and had a time complexity of $O(2n)$. This was generalized by Turau in \cite{TURAU200788} where he used an unfair distributed scheduler. Turau's algorithm had a time complexity of $max(3n - 5, 2n)$ moves. As part of the rule that handles when a node should go into the independent set, he has an $id(i)$ variable that is a unique numerical identifier of a node. This is similar to how we rank the algorithms by ids. For additional algorithms on maximal independent sets, \cite{cone2022performancecomparisonsselfstabilizingalgorithms}. compares 10 different versions for time complexity and performance in simulations.

An extension to Turau's algorithm, to account for the combined algorithms that we are proposing, is provided below. At convergence, there will be $X = \{ i$ $ |  x_{MIS}(i) = 1\}$ that forms a maximal independent set.

\begin{algorithm}[htbp]
 \begin{algorithmic}
    \State \textbf{RWait: if} x$_{MIS}$(i) = ${out}$ ${\wedge}$ (${\nexists}$j ${\in}$ N(i))(x$_{MIS}$(j) = ${in}$) ${\wedge}$ (${\nexists}$a ${\in}$ $ALG$)($a < MIS$ ${\wedge}$ x$_{a}$(i) = ${out}$)
    \State ${\;}$ \textbf{then} x$_{MIS}$(i) = ${wait}$

    \State \textbf{RBack: if} x$_{MIS}$(i) = ${wait}$ ${\wedge}$ ((${\exists}$j ${\in}$ N(i))(x$_{MIS}$(j) = ${in}$) ${\vee}$ (${\exists}$a ${\in}$ $ALG$)($a < MIS$ ${\wedge}$ x$_{a}$(i) = ${out}$))
    \State ${\;}$ \textbf{then} x$_{MIS}$(i) = ${out}$

    \State \textbf{RIn: if} x$_{MIS}$(i) = ${wait}$ ${\wedge}$ (${\nexists}$j ${\in}$ N(i))(x$_{MIS}$(j) = ${in}$) ${\wedge}$ (${\forall}$k ${\in}$ N(i))(x$_{MIS}$(k) ${\neq}$ ${wait}$ ${\vee}$ id(k) ${>}$ id(i)) ${\wedge}$ (${\nexists}$a ${\in}$ $ALG$)($a < MIS$ ${\wedge}$ x$_{a}$(i) = ${out}$)
    \State ${\;}$ \textbf{then} x$_{MIS}$(i) = ${in}$

    \State \textbf{ROut: if} x$_{MIS}$(i) = ${in}$ ${\wedge}$ ((${\exists}$j ${\in}$ N(i))(x$_{MIS}$(j) = ${in}$) ${\vee}$ (${\exists}$a ${\in}$ $ALG$)($a < MIS$ ${\wedge}$ x$_{a}$(i) = ${out}$)))
    \State ${\;}$ \textbf{then} x$_{MIS}$(i) = ${out}$
 \end{algorithmic}
 \caption{Maximal Independent Set}
 \label{alg:mis_combined}
\end{algorithm}

\lemma{For the subgraph induced by $(a \in ALG)(a < MIS)$, the correctness of the algorithm is not affected.}

\textbf{Proof:} The additional logic of determining if a node is in the $in/out$ set for an algorithm $(a < MIS)$ results in an induced subgraph, such that the maximal independent set will operate as previously on the set of nodes where $x_a(i) = in$. For nodes where $x_a(i) = out$, those will not be considered for the maximal independent set. \qed

\lemma{If there are no more moves for a node $i$ due to an algorithm $a$ in which $a < MIS$, the time complexity of the algorithm remains at most $max(3n - 5, 2n)$.}

\textbf{Proof:} For the time complexity to be increased from $max(3n - 5, 2n)$, this would require that an additional move is made due to the subgraph induced by $(a \in ALG)(a < MIS)$. However, as the subgraph is less than or the same size of nodes as $V$, it must be the case that the additional move(s) are due to the case of an algorithm $a < MIS$; however, this would also have been seen in lemmas 5-8 as both Algorithm~\ref{alg:bw_combined} and Algorithm~\ref{alg:mis_combined} share the same logic in that case. \qed

\theorem{Algorithm~\ref{alg:mis_combined} finds a maximal independent set in at most $max(3n - 5, 2n)$ moves on the subgraph induced by $(a \in ALG)(a < MIS)$.}

\textbf{Proof:} This follows directly from lemmas 9 and 10 and \cite{TURAU200788}. \qed

\subsection{Algorithm for 1-Minimal Dominating Set}\label{subsec:mds}

Hedetniemi et al. developed the first 1-minimal dominating set algorithm for self-stabilization in \cite{HEDETNIEMI2003805}. Their algorithm used a central scheduler and had a time complexity of $O(n^2)$. Various papers have improved upon this result; the one presented below is a modification from Chiu, Chen, and Tsai \cite{CHIU2014515}.

Chiu, Chen, and Tsai's algorithm uses an unfair distributed scheduler that has a time complexity of $O(4n)$. As with Turau, they used an $id(i)$ function that would return the unique numerical identifier of node $i$. As part of our modifications, $x(i)$ is a shared variable across the algorithms.

\begin{algorithm}[htbp]
 \begin{algorithmic}
    \State \textbf{RWait: if} (x$_{MDS}$(i) = ${out_1}$ ${\vee}$ x$_{MDS}$(i) = ${out_2}$) ${\wedge}$ (${\nexists}$j ${\in}$ N(i))(x$_{MDS}$(j) = ${in}$) ${\wedge}$ (${\nexists}$a ${\in}$ $ALG$)($a < MDS$ ${\wedge}$ x$_{a}$(i) = ${out}$)
    \State ${\;}$ \textbf{then} x$_{MDS}$(i) = ${wait}$

    \State \textbf{RBack$_1$: if} x$_{MDS}$(i) = ${wait}$ ${\wedge}$ ((${\vert}\{$j ${\in}$ N(i) ${\vert}$ x$_{MDS}$(j) = ${in}\}{\vert}$ = 1) ${\vee}$ (${\exists}$a ${\in}$ $ALG$)($a < MDS$ ${\wedge}$ x$_{a}$(i) = ${out}$)))
    \State ${\;}$ \textbf{then} x$_{MDS}$(i) = ${out_1}$

    \State \textbf{RBack$_2$: if} (x$_{MDS}$(i) = ${out_1}$ ${\vee}$ x$_{MDS}$(i) = ${wait}$) ${\wedge}$ ((${\vert}\{$j ${\in}$ N(i) ${\vert}$ x$_{MDS}$(j) = ${in}\}{\vert}$ ${>}$ 1) ${\vee}$ (${\exists}$a ${\in}$ $ALG$)($a < MDS$ ${\wedge}$ x$_{a}$(i) = ${out}$)))
    \State ${\;}$ \textbf{then} x$_{MDS}$(i) = ${out_2}$
    
    \State \textbf{RIn: if} x$_{MDS}$(i) = ${wait}$ ${\wedge}$ (${\nexists}$j ${\in}$ N(i))(x$_{MDS}$(j) = ${in}$) ${\wedge}$ (${\nexists}$k ${\in}$ N(i))(x$_{MDS}$(k) = ${wait}$ ${\wedge}$ id(k) ${<}$ id(i)) ${\wedge}$ (${\nexists}$a ${\in}$ $ALG$)($a < MDS$ ${\wedge}$ x$_{a}$(i) = ${out}$)
    \State ${\;}$ \textbf{then} x$_{MDS}$(i) = ${in}$

    \State \textbf{ROut$_1$: if} x$_{MDS}$(i) = ${in}$ ${\wedge}$ ((${\vert}\{$j ${\in}$ N(i) ${\vert}$ x$_{MDS}$(j) = ${in}\}{\vert}$ ${=}$ 1) ${\wedge}$ (${\nexists}$k ${\in}$ N(i))(x$_{MDS}$(k) = ${out_1}$) ${\vee}$ (${\exists}$a ${\in}$ $ALG$)($a < MDS$ ${\wedge}$ x$_{a}$(i) = ${out}$)))
    \State ${\;}$ \textbf{then} x$_{MDS}$(i) = ${out_1}$

    \State \textbf{ROut$_2$: if} x$_{MDS}$(i) = ${in}$ ${\wedge}$ ((${\vert}\{$j ${\in}$ N(i) ${\vert}$ x$_{MDS}$(j) = ${in}\}{\vert}$ ${>}$ 1) ${\wedge}$ (${\nexists}$k ${\in}$ N(i))(x$_{MDS}$(k) = ${out_1}$) ${\vee}$ (${\exists}$a ${\in}$ $ALG$)($a < MDS$ ${\wedge}$ x$_{a}$(i) = ${out}$)))
    \State ${\;}$ \textbf{then} x$_{MDS}$(i) = ${out_2}$
 \end{algorithmic}
 \caption{1-Minimal Dominating Set}
 \label{alg:mds_combined}
\end{algorithm}

\lemma{For the subgraph induced by $(a \in ALG)(a < MDS)$, the correctness of the algorithm is not affected.}

\textbf{Proof:} This follows from lemma 9. \qed

\lemma{For the subgraph induced by $(a \in ALG)(a < MDS)$, the time complexity of the algorithm is not affected.}

\textbf{Proof:} This follows from lemma 10. \qed

\theorem{Algorithm~\ref{alg:mds_combined} finds a 1-minimal dominating set in at most $O(4n)$ moves on the subgraph induced by $(a \in ALG)(a < MDS)$.}

\textbf{Proof:} This follows directly from lemmas 11, 12 and \cite{CHIU2014515}. \qed

\subsection{Correctness of Combined Algorithms}

The correctness of the individual algorithms, for the subgraphs that were induced by algorithms with a lower id, was shown in \ref{subsec:bw}, \ref{subsec:mis}, and \ref{subsec:mds}.

The final result will be a set $X$ of nodes, which will have values for each of the algorithms. Note that since the algorithms have a hierarchical order, it is possible to have a scenario such as the following: $BW < MIS$ with $x_{BW}(i) = in$ and $x_{MIS}(i) = out$. As the algorithms only determine if a node should be in/out of the set from their current state and the processing of algorithms with a lower id, algorithms with a higher id might result in subsequent removal of a node. 

Obviously for the most restrictive IP protection, the results of all of the algorithms need to be examined for each node with only the set ${in}$ included. However for complex graphs, there may not be a solution available when considering all dimensions. In these cases, it will have to be determined what risks are manageable.

Therefore, the most important IP dimensions should be provided the lowest ids so that solutions can be provided for these dimensions. Then the risk of subsequent dimensions can be considered.

\subsection{Time Complexity of Combined Algorithms}

The time complexity of the algorithms is shown in Table~\ref{tab:complexity}. These can also be found in the algorithm discussions. As the algorithms are hierarchical due to using ids, their combined worst case running time thus becomes:

\begin{table}
    \centering
    \begin{tabular}{c|c}
        \multicolumn{1}{p{2cm}|}{\centering \textbf{Algorithm}} & \multicolumn{1}{|p{2cm}}{\centering \textbf{Time Complexity}} \\
        \hline
        Whitelists and Blacklists & $O(2n)$\\
        \hline
        Maximal Independent Set & $max(3n - 5, 2n)$\\
        \hline
        1-Minimal Dominating Set & $O(4n)$\\
        \hline
    \end{tabular}
    \caption{Time complexity of presented algorithms.}
    \label{tab:complexity}
\end{table}

\begin{equation}
    O(a_1) + O(a_1)*O(a_2) + O(a_1)*O(a_2)*O(a_3) + ....
\end{equation}

This is represented by the following product summation:

\begin{equation}
    \sum_{k=1}^{|ALG|} \prod_{i=1}^{k} O(a_i)
    \label{eq:combined_algs}
\end{equation}

The complexity is thus polynomial - with the specific value dependent upon the algorithms that are used and their hierarchical order. For the use case that we have been describing in this paper, we have the following algorithm order:

\begin{itemize}
    \item $a_1$: white/blacklists
    \item $a_2$: maximal independent set for suppliers
    \item $a_3$: 1-minimal dominating set for parts flow
\end{itemize}

The worst case combined time complexity is then:

\begin{equation}
    O(24n^3 - 34n^2 - 8n) = O(24n^3)
    \label{eq:time_complexity}
\end{equation}

\section{Conclusions and Further Work}\label{sec:conclusions}

In this paper, we developed a solution of polynomial time complexity as opposed to the exponential time complexity in \cite{7139023}. This allows for us to use the proposed method at scale for real-world problems in which we are interested in determining intellectual property risk when examining various dimensions.

Additionally, we showed a different number of algorithms that could be used for a variety of different IP risk management scenarios  (e.g. parts flow, protecting specific areas, ensuring that all information regarding associates/equipment/suppliers not made available, ...).

Future work could include the use of other algorithms (e.g. graph coloring, bipartite, ...) that would be able to handle further types of IP dimensions.

\bibliographystyle{IEEEtran}
\bibliography{references}

@article{10.1142/S0218488502001648,
    author = {Sweeney, Latanya},
    title = {k-anonymity: a model for protecting privacy},
    year = {2002},
    issue_date = {October 2002},
    publisher = {World Scientific Publishing Co., Inc.},
    address = {USA},
    volume = {10},
    number = {5},
    issn = {0218-4885},
    url = {https://doi.org/10.1142/S0218488502001648},
    doi = {10.1142/S0218488502001648},
    journal = {Int. J. Uncertain. Fuzziness Knowl.-Based Syst.},
    month = oct,
    pages = {557–570},
    numpages = {14}
}

@article{10.1145/1217299.1217302,
    author = {Machanavajjhala, Ashwin and Kifer, Daniel and Gehrke, Johannes and Venkitasubramaniam, Muthuramakrishnan},
    title = {L-diversity: Privacy beyond k-anonymity},
    year = {2007},
    issue_date = {March 2007},
    publisher = {Association for Computing Machinery},
    address = {New York, NY, USA},
    volume = {1},
    number = {1},
    issn = {1556-4681},
    url = {https://doi.org/10.1145/1217299.1217302},
    doi = {10.1145/1217299.1217302},
    journal = {ACM Trans. Knowl. Discov. Data},
    month = mar,
    pages = {3–es},
    numpages = {52}
}

@INPROCEEDINGS{4221659,
    author={Li, Ninghui and Li, Tiancheng and Venkatasubramanian, Suresh},
    booktitle={2007 IEEE 23rd International Conference on Data Engineering}, 
    title={t-Closeness: Privacy Beyond k-Anonymity and l-Diversity}, 
    year={2007},
    volume={},
    number={},
    pages={106-115},
    doi={10.1109/ICDE.2007.367856}
}

@INPROCEEDINGS{7004228,
    author={Anderson, Jason W. and Kennedy, K. E. and Ngo, Linh B. and Luckow, Andre and Apon, Amy W.},
    booktitle={2014 IEEE International Conference on Big Data (Big Data)}, 
    title={Synthetic data generation for the internet of things}, 
    year={2014},
    volume={},
    number={},
    pages={171-176},
    doi={10.1109/BigData.2014.7004228}
}

@InProceedings{10.1007/11787006_1,
    author="Dwork, Cynthia",
    editor="Bugliesi, Michele
    and Preneel, Bart
    and Sassone, Vladimiro
    and Wegener, Ingo",
    title="Differential Privacy",
    booktitle="Automata, Languages and Programming",
    year="2006",
    publisher="Springer Berlin Heidelberg",
    address="Berlin, Heidelberg",
    pages="1--12",
    isbn="978-3-540-35908-1"
}

@Inbook{Raskhodnikova2014,
    author="Raskhodnikova, Sofya and Smith, Adam",
    editor="Kao, Ming-Yang",
    title="Private Analysis of Graph Data",
    bookTitle="Encyclopedia of Algorithms",
    year="2014",
    publisher="Springer Berlin Heidelberg",
    address="Berlin, Heidelberg",
    pages="1--6",
    isbn="978-3-642-27848-8"
}

@Inbook{Raskhodnikova2016,
    author="Raskhodnikova, Sofya and Smith, Adam",
    editor="Kao, Ming-Yang",
    title="Differentially Private Analysis of Graphs",
    bookTitle="Encyclopedia of Algorithms",
    year="2016",
    publisher="Springer New York",
    address="New York, NY",
    pages="543--547",
    isbn="978-1-4939-2864-4"
}

@article{DBLP:journals/corr/abs-2010-08688,
    author       = {Jacob Imola and
                  Takao Murakami and
                  Kamalika Chaudhuri},
    title        = {Locally Differentially Private Analysis of Graph Statistics},
    journal      = {CoRR},
    volume       = {abs/2010.08688},
    year         = {2020},
    url          = {https://arxiv.org/abs/2010.08688},
    eprinttype    = {arXiv},
    eprint       = {2010.08688},
    bibsource    = {dblp computer science bibliography, https://dblp.org}
}

@article{DBLP:journals/corr/abs-1204-2136,
    author       = {Jeremiah Blocki and
                  Avrim Blum and
                  Anupam Datta and
                  Or Sheffet},
    title        = {The Johnson-Lindenstrauss Transform Itself Preserves Differential
                  Privacy},
    journal      = {CoRR},
    volume       = {abs/1204.2136},
    year         = {2012},
    url          = {http://arxiv.org/abs/1204.2136},
    eprinttype    = {arXiv},
    eprint       = {1204.2136},
    bibsource    = {dblp computer science bibliography, https://dblp.org}
}

@inproceedings{10.1007-978-3-642-36594-2-26,
    author = {Kasiviswanathan, Shiva Prasad and Nissim, Kobbi and Raskhodnikova, Sofya and Smith, Adam},
    title = {Analyzing graphs with node differential privacy},
    year = {2013},
    isbn = {9783642365935},
    publisher = {Springer-Verlag},
    address = {Berlin, Heidelberg},
    booktitle = {Proceedings of the 10th Theory of Cryptography Conference on Theory of Cryptography},
    pages = {457–476},
    numpages = {20},
    location = {Tokyo, Japan},
    series = {TCC'13}
}

@INPROCEEDINGS{7139023,
    author={O'Kane, Jason M. and Shell, Dylan A.},
    booktitle={2015 IEEE International Conference on Robotics and Automation (ICRA)}, 
    title={Automatic design of discreet discrete filters}, 
    year={2015},
    volume={},
    number={},
    pages={353-360},
    doi={10.1109/ICRA.2015.7139023}
}

@InProceedings{10.1007-978-3-030-44051-0-7,
    author="Zhang, Yulin
    and Shell, Dylan A.
    and O'Kane, Jason M.",
    editor="Morales, Marco
    and Tapia, Lydia
    and S{\'a}nchez-Ante, Gildardo
    and Hutchinson, Seth",
    title="Finding Plans Subject to Stipulations on What Information They Divulge",
    booktitle="Algorithmic Foundations of Robotics XIII",
    year="2020",
    publisher="Springer International Publishing",
    address="Cham",
    pages="106--124",
    isbn="978-3-030-44051-0"
}

@article{10.1155-2018-3858592,
    author = {Zare-Garizy, Tirazheh and Fridgen, Gilbert and Wederhake, Lars},
    title = {A Privacy Preserving Approach to Collaborative Systemic Risk Identification: The Use-Case of Supply Chain Networks},
    journal = {Security and Communication Networks},
    volume = {2018},
    number = {1},
    pages = {3858592},
    year = {2018}
}

@book{haynes2013fundamentals,
    title={Fundamentals of domination in graphs},
    author={Haynes, Teresa W and Hedetniemi, Stephen and Slater, Peter},
    year={2013},
    publisher={CRC press}
}

@article{dijkstra1974self,
    title={Self-stabilizing systems in spite of distributed control},
    author={Dijkstra, Edsger W},
    journal={Communications of the ACM},
    volume={17},
    number={11},
    pages={643--644},
    year={1974},
    publisher={ACM New York, NY, USA}
}

@article{dijkstra1985belated,
    title={A belated proof of self-stabilization},
    author={Dijkstra, Edsger W},
    journal={Distrib Comput},
    volume={1},
    pages={5-6},
    year={1986}
}

@misc{neumann1966theory,
    title={Theory of self-reproducing automata},
    author={Neumann, John von and Burks, Arthur W and others},
    year={1966},
    publisher={University of Illinois press Urbana}
}

@article{cockayne1987k,
    title={k-Minimal domination numbers of cycles},
    author={Cockayne, EJ and Mynhardt, CM},
    journal={Ars Combinatoria},
    volume={23},
    pages={195--206},
    year={1987}
}

@article{shukla,
    author = {Shukla, Sandeep and Rosenkrantz, Daniel and Ravi, S.},
    year = {1995},
    month = {01},
    pages = {},
    title = {Observations on self-stabilizing graph algorithms for anonymous networks},
    journal = {Proceedings of the Second Workshop on Self-Stabilizing Systems}
}

@article{HEDETNIEMI2003805,
    title = {Self-stabilizing algorithms for minimal dominating sets and maximal independent sets},
    journal = {Computers \& Mathematics with Applications},
    volume = {46},
    number = {5},
    pages = {805-811},
    year = {2003},
    issn = {0898-1221},
    doi = {https://doi.org/10.1016/S0898-1221(03)90143-X},
    url = {https://www.sciencedirect.com/science/article/pii/S089812210390143X},
    author = {S.M. Hedetniemi and S.T. Hedetniemi and D.P. Jacobs and P.K. Srimani}
}

@article{CHIU2014515,
    title = {A 4n-move self-stabilizing algorithm for the minimal dominating set problem using an unfair distributed daemon},
    journal = {Information Processing Letters},
    volume = {114},
    number = {10},
    pages = {515-518},
    year = {2014},
    issn = {0020-0190},
    doi = {https://doi.org/10.1016/j.ipl.2014.04.011},
    url = {https://www.sciencedirect.com/science/article/pii/S0020019014000702},
    author = {Well Y. Chiu and Chiuyuan Chen and Shih-Yu Tsai}
}

@article{TURAU200788,
    title = {Linear self-stabilizing algorithms for the independent and dominating set problems using an unfair distributed scheduler},
    journal = {Information Processing Letters},
    volume = {103},
    number = {3},
    pages = {88-93},
    year = {2007},
    issn = {0020-0190},
    doi = {https://doi.org/10.1016/j.ipl.2007.02.013},
    url = {https://www.sciencedirect.com/science/article/pii/S0020019007000488},
    author = {Volker Turau}
}

@misc{cone2022performancecomparisonsselfstabilizingalgorithms,
    title={Performance Comparisons of Self-stabilizing Algorithms for Maximal Independent Sets}, 
    author={Barton F. Cone and Stephen T. Hedetniemi and Lance C. Ingle and Ken Kennedy},
    year={2022},
    eprint={2203.11118},
    archivePrefix={arXiv},
    primaryClass={cs.DC},
    url={https://arxiv.org/abs/2203.11118}
}

\end{document}